\documentclass[11pt,twoside]{article}
\usepackage{asp2010}

\begin{document}

\title{An ADER-WENO Finite Volume AMR code for Astrophysics} 
\author{O.~Zanotti$^1$, M.~Dumbser$^1$, A.~Hidalgo$^2$ and D.~Balsara$^3$
\affil{$^1$Laboratory of Applied Mathematics, Department
  of Civil, Environmental and Mechanical Engineering,
  University of Trento, Via Mesiano 77, I-38123 Trento,
  Italy}
\affil{$^2$Departamento de Matem\'atica Aplicada y M\'etodos Inform\'aticos, Universidad Polit\'ecnica de Madrid, Calle R\'ios Rosas 21, E-28003 Madrid, Spain}
\affil{$^3$Physics Department, University of Notre Dame du Lac, 225 Nieuwland Science Hall, Notre Dame, IN 46556, USA}
}

\begin{abstract}
A high order one-step ADER-WENO finite
volume scheme with Adaptive Mesh Refinement (AMR) in
multiple space  dimensions is presented.
A high order one-step time
discretization 
is achieved using a local space-time discontinuous
Galerkin predictor method, while 
a high order spatial accuracy is obtained through a WENO
reconstruction.
Thanks to the one-step nature of
the underlying scheme, the resulting algorithm 
can be efficiently imported within an 
AMR framework on
space-time adaptive meshes.
We provide convincing evidence that
the presented high order AMR scheme
behaves better  than traditional second order AMR
methods. 
Tests are shown of the 
new scheme for
nonlinear systems of hyperbolic conservation laws,
including the  classical Euler equations 
and the equations of ideal magnetohydrodynamics.
The proposed scheme  is likely to
become a useful tool in several astrophysical scenarios.
\end{abstract}

\section{Introduction}
\label{generalintroduction}
ADER schemes were introduced more than ten years ago by 
\citet{toro1}\footnote{See also the subsequent works by  
\citet{titarevtoro, titarevtoro2}.}. Though presenting
several attractive features, like the possibility of
reaching an arbitrary high order of accuracy and the fact
that the time update is performed through a single step
with no need for Runge-Kutta, in their original form they have never been
applied to astrophysics.
The reason for this is that
original ADER schemes required 
resorting to the so-called 
Cauchy-Kovalewski procedure, that is based on a repeated use 
of the governing conservation laws in differential form.
In spite of being conceptually
clear, such a procedure becomes rather cumbersome in
multiple space dimensions for the classical Euler
equations. It has been implemented by 
\citet{Toro2005} and \citet{DumbserKaeser07}, but 
a wider use has not been reached. 

A fundamental breakthrough was achieved by
\citet{DumbserEnauxToro}, who proposed a variant of the
ADER approach in which the local time evolution of the 
reconstructed polynomials is obtained through an
element-local  space-time Galerkin predictor that is
based on the weak integral form of the PDEs. This new
strategy, which promoted ADER methods to the wider public,
 has been proved to be rather successful even in
the solution of systems of equations with stiff source
terms, and has been applied to a variety of 
specific physical problems
\citep{DumbserZanotti, HidalgoDumbser, ZanottiDumbser2011}.

The ADER property of performing the time-update through a
single step makes them particularly suitable for
incorporating adaptive mesh refinement (AMR). This combination 
is the subject of our work, which describes
a new tool for performing high-order numerical
calculations on adaptive grids. We refer to
\cite{Dumbser2013} for more details about the new
approach, and to
\citet{Dumbser2014} for the application of the same
ideas to non-conservative hyperbolic systems.

\section{The numerical scheme}
The partial differential equations under consideration
are of the type
\begin{equation}
\label{eq:governingPDE}
\frac{\partial{\bf u}}{\partial t}+\frac{\partial{\bf f}}{\partial x}+\frac{\partial{\bf g}}{\partial y}+\frac{\partial{\bf h}}{\partial z}={\bf S}({\bf u},\mathbf{x},t)\,,
\end{equation}
where $\bf {u}$ is the vector of conserved quantities,
while ${\bf f}(\bf{u})$, ${\bf g}(\bf{u})$ and ${\bf
  h}(\bf{u})$ are the  fluxes. According to the finite
volume approach, the time update is performed through the
standard discretization
\begin{eqnarray}
\label{eq:finite_vol}
{\bf \bar u}_{ijk}^{n+1}&=&{\bf \bar u}_{ijk}^{n}-\frac{\Delta t}{\Delta x_i}\left({\bf f}_{i+{\frac{1}{2}},jk}-{\bf f}_{i-{\frac{1}{2}},jk} \right)-\frac{\Delta t}{\Delta y_j}\left({\bf g}_{i,j+{\frac{1}{2}},k}-{\bf g}_{i,j-{\frac{1}{2}},k} \right)
\nonumber \\
&& \hspace{8mm} -\frac{\Delta t}{\Delta z_k}\left({\bf h}_{ij,k+{\frac{1}{2}}}-{\bf h}_{ij,k-{\frac{1}{2}}} \right)+ \Delta t {\bf \bar{S}}_{ijk}\,.
\end{eqnarray}
In a nutshell, the modern ADER strategy provides 
an implementation of Eq.~(\ref{eq:finite_vol}) through the
following three
steps
\begin{itemize}
\item {\bf High order reconstruction.} This is performed
  through WENO schemes, in which the sought reconstructed
  polynomial is written in terms of a nodal basis
  $\psi_p(\xi)$ of
  polynomials of degree $M$. In one spatial dimension, say $x$, this
  corresponds to the expansion
\begin{equation}
\label{eqn.recpolydef.x} 
 \mathbf{w}^{s,x}_h(x,t^n) = \sum \limits_{p=0}^M \psi_p(\xi) \hat \mathbf{w}^{n,s}_{ijk,p}\,,
\end{equation}
where $\xi\in[0,1]$ is the reference coordinate
defined by $x = x_{i-1/2} + \xi   \Delta x_i$,
$\mathbf{w}^{n,s}_{ijk,p}$ are unknown coefficients
to be determined, 
while the index $s$ refers to the
stencil for which reconstruction is performed [see
\citet{DumbserEnauxToro} for more details].

\item {\bf Local Predictor in time.} Once a high order
  polynomial in space $\mathbf{w}_h$ has been
  reconstructed for each cell, it is necessary to evolve
  it in time in order  to compute the fluxes while
  preserving the order of accuracy.
This is
obtained after writing a local weak form of the governing
equations, namely
\begin{equation}
 \int \limits_{0}^{1} \int \limits_{0}^{1}  \int \limits_{0}^{1}   \int \limits_{0}^{1}   
\theta_\mathfrak{q} \left( 
  \frac{\partial{\bf u}}{\partial \tau} + \frac{\partial \mathbf{f}^\ast}{\partial \xi} + \frac{\partial \mathbf{g}^\ast}{\partial \eta} + \frac{\partial \mathbf{h}^\ast}{\partial \zeta} - {\bf S}^\ast \right) d\xi d\eta d\zeta d\tau = 0\,,  
\label{eqn.pde.weak1} 
\end{equation}
where the $\theta_\mathfrak{p}$ are given by a
tensor--product of the basis functions  $\psi_p$ already
used in the reconstruction procedure, while 
\begin{equation}
{\bf f}^\ast= \frac{\Delta t}{\Delta x_i} \, {\bf f}, \quad 
{\bf g}^\ast= \frac{\Delta t}{\Delta y_j} \, {\bf g}, \quad 
{\bf h}^\ast= \frac{\Delta t}{\Delta z_k} \, {\bf h}, \quad 
{\bf S}^\ast= \Delta t {\bf S}. 
\end{equation}
are the fluxes and sources over the reference
coordinates. After integration by parts in time,
Eq.~(\ref{eqn.pde.weak1}) leads to the solution of a
local nonlinear system of equations in the unknowns
degrees of freedom of the reconstructed polynomials.

\item {\bf Time update.} This is essentially obtained
  through the scheme (\ref{eq:finite_vol}), in which 
  the fluxes can be computed with a high order of
  accuracy, using the information obtained from the local predictor.
\end{itemize}

\section{Adaptive Mesh Refinement}
Adaptive mesh refinement has been implemented according
to a ``cell-by-cell''  approach involving a tree-type data
structure, which is similar to that of \cite{Khokhlov1998}.
An arbitrary number of levels of refinement
$\ell$ and of the refinement factor $\mathfrak{r}$
can in principle be used, 
where
\begin{equation}
\label{refine-factor}
\Delta x_{\ell} = \mathfrak{r} \Delta x_{\ell+1}\, \quad \Delta y_{\ell} = \mathfrak{r} \Delta y_{\ell+1} \, \quad 
\Delta z_{\ell} = \mathfrak{r} \Delta z_{\ell+1}\,.
\end{equation}
In practice,
we have typically
used $\ell_{\max}=2$ and $\mathfrak{r}=3$.
The computation of numerical fluxes between two adjacent cells on different levels of 
refinement is rather straightforward thanks to the use of the local space--time predictor, 
which computes the predictor solution 
for each element after reconstruction and 
which is valid from time $t^n_\ell$ to time
$t^{n+1}_\ell$.  We emphasize that local time-stepping is
naturally obtained, implying 
a total amount of $\mathfrak{r}^\ell$ sub-timesteps on 
each level to be performed, in order to reach the time
$t_0^{n+1}$ of the coarsest level.  

\section{Results}

\begin{figure}
\begin{center}
\includegraphics[width=0.95\textwidth]{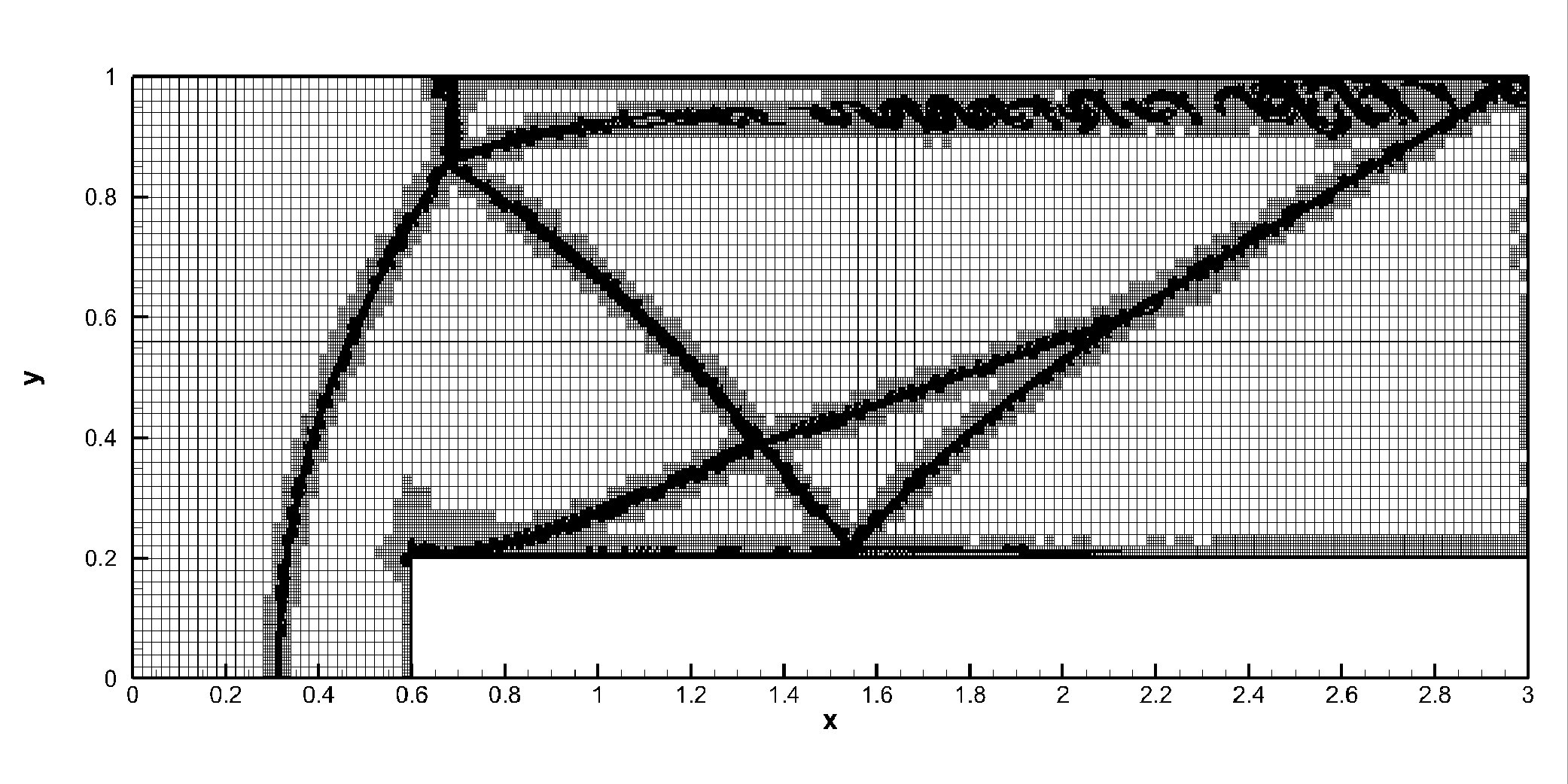} \\ 
\includegraphics[width=0.95\textwidth]{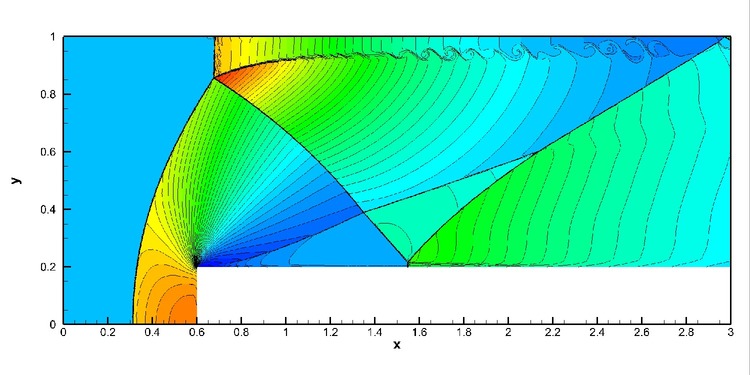} \\  
\includegraphics[width=0.95\textwidth]{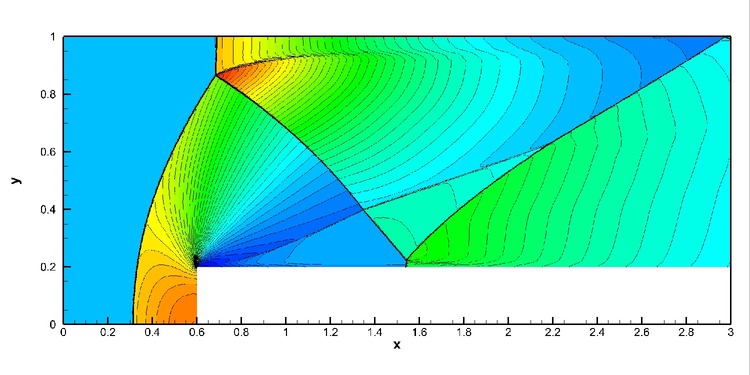}    
\caption{ Forward facing step problem. Top panel:
AMR grid. Central panel: density contours obtained  
with the third order ADER-WENO scheme, at time $t=2.5$. 
Bottom panel: density contours obtained  
with the second order ADER-WENO scheme.
Reproduced with courtesy from Journal of Computational
Physics, \cite{Dumbser2013}. 
}
\label{fig.ffs}
\end{center}
\end{figure}

We have validated our scheme through a large set of test
problems in all spatial dimensions, by solving the
classical hydrodynamics and magnetohydrodynamics equations.
A limited sample of such tests is reported
here. 

Figure~\ref{fig.ffs} shows the result of the {\em
  forward facing step} problem by \citet{woodwardcol84},
which has been solved twice, once at the third order
of accuracy and once at the second order,
while using the same numerical grid.
The grid on the coarsest level is formed by 
$150 \times 50$ control volumes, with
$\mathfrak{r}=4$ and $\ell_{\max}=2$.
The top panel shows the refined grid at time $t=2.5$;
the central panel
refers to the third-order ADER-WENO computation, while the
bottom panel refers to the second order computation.
This  indicates that even in the context of space--time
adaptive mesh refinement, the use of higher order schemes
can become crucial to highlight the formation of
small-scale turbulent structures.      

\begin{figure}
\begin{center}
\begin{tabular}{lcr}
\includegraphics[width=0.45\textwidth]{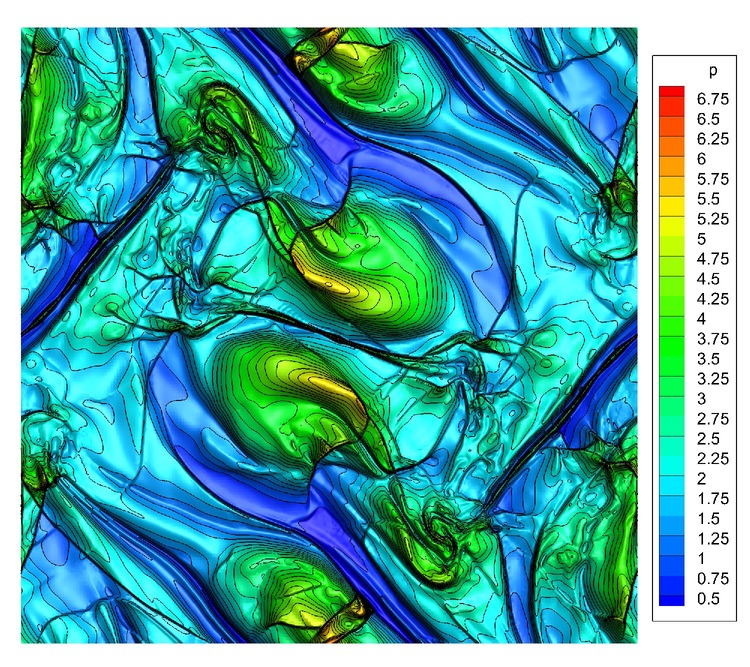}      &  
\includegraphics[width=0.45\textwidth]{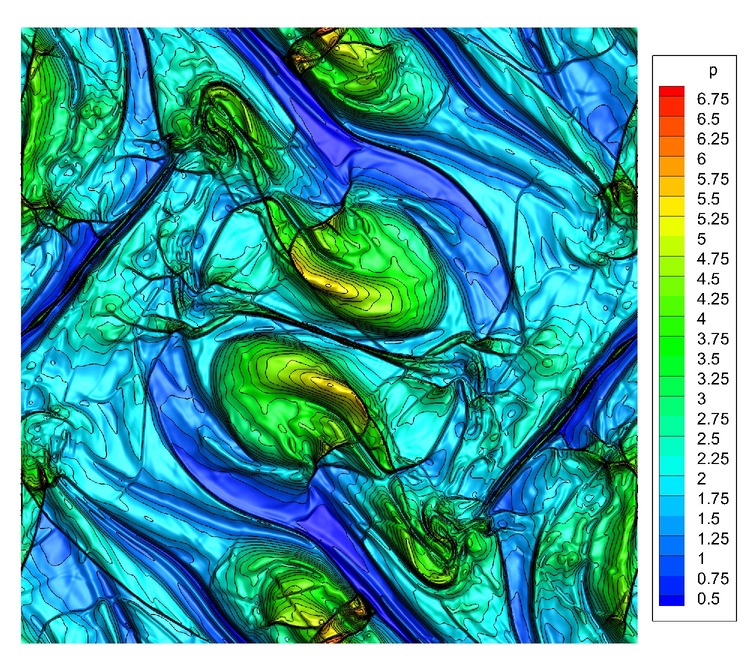}      
\end{tabular} 
\caption{ Orszag-Tang vortex system at time $t=5.0$. Third order ADER-WENO solution obtained on the AMR grid (left) and on a fine uniform 
grid corresponding to the finest AMR grid level (right).
Reproduced with courtesy from Journal of Computational
Physics, \cite{Dumbser2013}.
} 
\label{fig.ot}
\end{center}
\end{figure}

Figure~\ref{fig.ot} shows the distribution of the pressure
in the 
classical {\em vortex system} 
of \citet{OrszagTang}, with initial
conditions given by
\begin{equation}
  \left(\rho,u,v,p,B_x,B_y\right) = \left( \gamma^2, -\sin(y), \sin(x), \gamma, -\sqrt{4\pi} \sin(y), \sqrt{4\pi} \sin(2x)  \right), 
\end{equation}
where $B_z=0$ and $\gamma = 5/3$. The AMR computation
is reported in the central panel, while the right panel
shows a computation obtained 
on a fine uniform 
grid that corresponds to the finest AMR grid level. The
agreement between the two is excellent, but the AMR
simulation,
even for such a test where most of the cells are refined,
guarantees a
speedup of a factor of 1.8 compared to the uniform fine
mesh simulation.

\begin{figure}
\begin{center}
\begin{tabular}{lr}
\includegraphics[width=0.45\textwidth]{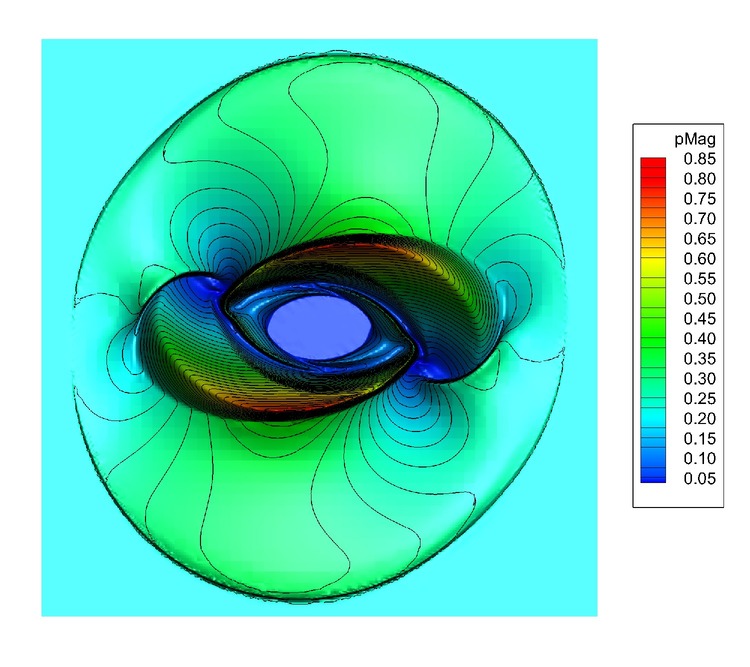}      &  
\includegraphics[width=0.45\textwidth]{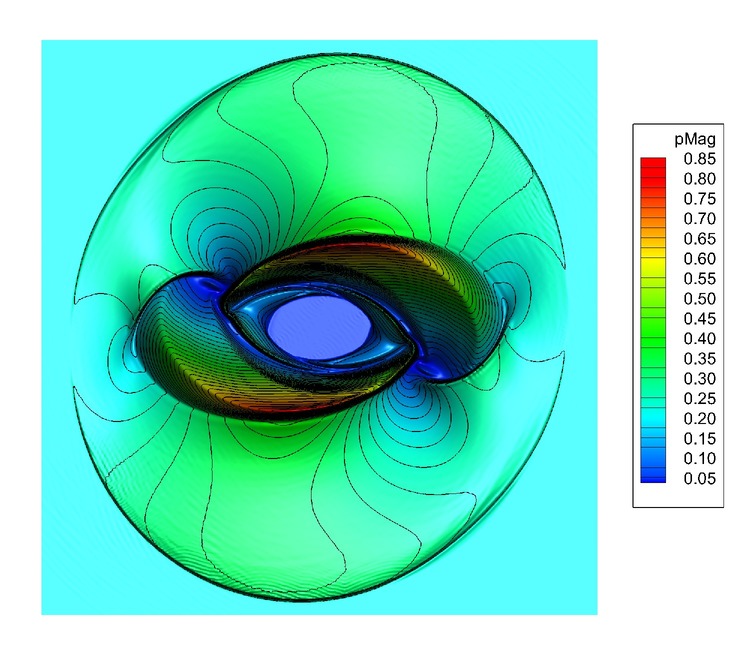}      
\end{tabular} 
\caption{ Magnetic pressure in the MHD rotor problem at time $t=0.25$. Third order ADER-WENO solution obtained on the AMR grid (left) and on a fine uniform 
grid corresponding to the finest AMR grid level (right).
Reproduced with courtesy from Journal of Computational
Physics, \cite{Dumbser2013}.
} 
\label{fig.mhdrotor}
\end{center}
\end{figure}

Figure~\ref{fig.mhdrotor}
shows the distribution of the magnetic pressure
in the {\em MHD rotor problem} in which a rapidly rotating
high density fluid is embedded in low density fluid at
rest~\citep{BalsaraSpicer1999}. 
On the level zero, the mesh contains $60 \times 60$
elements, with $\mathfrak{r}=4$ and $\ell_{\max}=2$.
Again, the AMR
computation is reported on the left, while the right
panel reports the result obtained with 
a fine uniform 
grid corresponding to the finest AMR grid level. In this
case, the benefit of the AMR strategy is particularly
evident:
while at the final time the 
AMR grid has only $179680$ elements, the uniform fine
grid has $921600$ elements. This has a strong impact on
the CPU time needed for performing the computation,
which is a factor $0.14$ smaller when the AMR method is adopted.

\section{Conclusions}
\label{sec:concl}

We have described the implementation of 
ADER-WENO methods
with AMR techniques. This combination is rather natural
within the ADER approach, which allows for numerical
schemes with a single step for the time-update. 
By providing high order of accuracy both in space and in
time, the new scheme is likely to give a significant
contribution to the study of delicate astrophysical
problems such as turbulence and fluid instabilities,
which will be the subject of future research.

\bibliography{aspzanotti}

\begin{thebibliography}{}
\expandafter\ifx\csname natexlab\endcsname\relax\def\natexlab#1{#1}\fi
\expandafter\ifx\csname url\endcsname\relax
  \def\url#1{\texttt{#1}}\fi
\expandafter\ifx\csname urlprefix\endcsname\relax\def\urlprefix{URL }\fi
\providecommand{\eprint}[2][]{\url{#2}}

\bibitem[{Balsara \& Spicer(1999)}]{BalsaraSpicer1999}
Balsara, D., \& Spicer, D. 1999, Journal of Computational Physics, 149, 270

\bibitem[{Dumbser et~al.(2008)Dumbser, Enaux, \& Toro}]{DumbserEnauxToro}
Dumbser, M., Enaux, C., \& Toro, E. 2008, Journal of Computational Physics,
  227, 3971

\bibitem[{Dumbser et~al.(2014)Dumbser, Hidalgo, \& Zanotti}]{Dumbser2014}
Dumbser, M., Hidalgo, A., \& Zanotti, O. 2014, Computer Methods in Applied
  Mechanics and Engineering, 268, 359-387

\bibitem[{Dumbser et~al.(2007)Dumbser, K\"aser, Titarev, \&
  Toro}]{DumbserKaeser07}
Dumbser, M., K\"aser, M., Titarev, V., \& Toro, E. 2007, Journal of
  Computational Physics, 226, 204

\bibitem[{Dumbser \& Zanotti(2009)}]{DumbserZanotti}
Dumbser, M., \& Zanotti, O. 2009, Journal of Computational Physics, 228, 6991

\bibitem[{{Dumbser} et~al.(2013){Dumbser}, {Zanotti}, {Hidalgo}, \&
  {Balsara}}]{Dumbser2013}
{Dumbser}, M., {Zanotti}, O., {Hidalgo}, A., \& {Balsara}, D.~S. 2013, Journal
  of Computational Physics, 248, 257

\bibitem[{Hidalgo \& Dumbser(2011)}]{HidalgoDumbser}
Hidalgo, A., \& Dumbser, M. 2011, Journal of Scientific Computing, 48, 173

\bibitem[{Khokhlov(1998)}]{Khokhlov1998}
Khokhlov, A. 1998, Journal of Computational Physics, 143, 519

\bibitem[{Orszag \& Tang(1979)}]{OrszagTang}
Orszag, S.~A., \& Tang, C.~M. 1979, Journal of Fluid Mechanics, 90, 129

\bibitem[{Titarev \& Toro(2005)}]{titarevtoro}
Titarev, V., \& Toro, E. 2005, Journal of Computational Physics, 204, 715

\bibitem[{Toro et~al.(2001)Toro, Millington, \& Nejad}]{toro1}
Toro, E., Millington, R., \& Nejad, L. 2001, in Godunov Methods. Theory and
  Applications, edited by E.~Toro (Kluwer/Plenum Academic Publishers), 905--938

\bibitem[{Toro \& Titarev(2005)}]{titarevtoro2}
Toro, E., \& Titarev, V. 2005, Journal of Computational Physics, 202, 196

\bibitem[{{Toro} \& {Titarev}(2005)}]{Toro2005}
{Toro}, E.~F., \& {Titarev}, V.~A. 2005, Journal of Computational Physics, 202,
  196

\bibitem[{Woodward \& Colella(1984)}]{woodwardcol84}
Woodward, P., \& Colella, P. 1984, Journal of Computational Physics, 54, 115

\bibitem[{Zanotti \& Dumbser(2011)}]{ZanottiDumbser2011}
Zanotti, O., \& Dumbser, M. 2011, Monthly Notices of the Royal Astronomical
  Society, 418, 1004

\end{thebibliography}
\bibliographystyle{asp2010}

\end{document}